\documentstyle[twoside,fleqn,espcrc2]{article}

\input epsf

\newcommand{\AmS}{{\protect\the\textfont2
  A\kern-.1667em\lower.5ex\hbox{M}\kern-.125emS}}

\title{Monopoles and deconfinement transition 
      in finite temperature $SU(2)$ QCD}
\author{Shun-ichi Kitahara\address{
        Department of Physics, Kanazawa University, 
        Kanazawa 920-11, Japan, 
        },
        Yoshimi Matsubara\address{Nanao Junior College, Nanao, 
        Ishikawa 926, Japan
        }%
        and 
        Tsuneo Suzuki\addtocounter{address}{-2}\addressmark  
}
\begin{document}

\begin{abstract}
We investigate the role of monopoles in the deconfinement transition 
of finite temperature $SU(2)$ QCD in the maximally abelian gauge.
In the confinement phase a long monopole loop exists in each configuration, 
whereas no long loop exists in the deep deconfinement region.
Balancing of the energy and the entropy of loops 
of the maximally extended monopoles can explain the occurrence of 
the deconfinement transition.
\end{abstract}

\maketitle

\section{Introduction}

Recent studies in abelian projected QCD on lattices show 
that condensation of monopoles seems to cause 
confinement of quarks \cite{suzu93}.
The 'tHooft's idea of abelian projection is to fix the gauge 
in such a way that the maximal torus group remains unbroken. 
QCD is regarded as an abelian theory with monopoles \cite{thooft2}.
If the monopoles make Bose condensation, 
quarks are confined due to the dual Meissner effect.

The long range property is expected to be important in the confinement 
phase of QCD.
Considering extended monopoles which corresponds to performing 
block spin transformations on the dual lattice 
is important \cite{ivanenko}.
Recently interesting results have been found \cite{shiba6}:
\vspace{-7pt}
\begin{enumerate}
\item A monopole effective action can be calculated for extended monopoles.
%% \vspace{-7pt}
%% \item To investigate the long range property of QCD, 
%%       considering extended monopoles which correspond to performing 
%%       block spin transformations on the dual lattice is important.
\vspace{-7pt}
\item As the extendedness of monopoles becomes bigger, $\beta_c$ tends 
      to be larger.
      Here $\beta_c$ is where balancing of the energy and 
      the entropy of a monopole loop occurs.
\vspace{-7pt}
\item The effective action for $n^3$ extended monopoles becomes the same 
      for any $n$ written in terms of $b$, $b=na(\beta)$.
      Hence corresponding to $\beta_c$, there exists a unique value 
      $b_c=5.2\times 10^{-3}\Lambda_L^{-1}$ for extended monopoles, 
      above which monopoles condense.
\vspace{-7pt}
\item Suppose that the above facts remain correct 
      in the infinite volume limit.
      QCD is always (for any $\beta$) in the monopole-condensed phase.
      Abelian charges after abelian projection are always confined and, 
      at the same time, color charges are confined.
\vspace{-3pt}
\end{enumerate}
What about $T\neq 0$ QCD?
The biggest difference from $T=0$ QCD is that the time extent is finite.
Because $N_t$ is finite, infinitely extended monopoles can not be adopted.
There is a maximum size of extendedness, so that 
$\beta_c$ is necessarily finite in contrast to the $T=0$ QCD case.
We investigate what are the maximally extended monopoles and 
whether $\beta_c$ can be explained by balancing of energy and entropy 
of the monopole loops.

\section{The role of monopoles in the deconfinement transition}

It has been found that the monopole action in the confinement phase 
in the $T\neq 0$ QCD is the same as that given in $T=0$ QCD \cite{kita}.
The energy of a monopole loop can be estimated 
by the self-energy term of the action.
However the entropy of the monopole loops must be affected 
by the finiteness of the time direction.
To study the entropy, we investigate the behavior of monopole loops 
in $T\neq 0$ $SU(2)$ QCD.
First, we calculate the length of monopole loops \cite{bode1}.
%
%---------- fig1 ---------
\begin{figure}[htb]
%\vspace{3pt}
%\framebox[55mm]{\rule[-21mm]{0mm}{43mm}}
\epsfxsize=0.5\textwidth
\vspace{-27pt}
\begin{center}
\leavevmode
\epsfbox{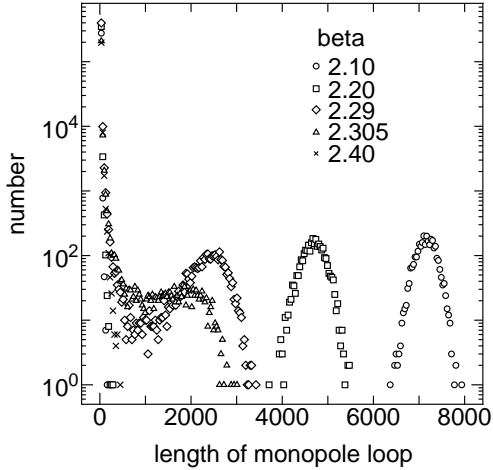}
\end{center}
\vspace{-50pt}
\caption{
The histogram of the length of $1^3$ monopole loops 
on $16^3\times 4$ lattices.
}
\label{fig:histogram}
\vspace{-15pt}
\end{figure}
%-------------------------
%
Fig.\ref{fig:histogram} shows the histogram of the length of monopole loops.
These data are obtained from 3000 configurations in the confinement region 
and 1500 configurations in the deconfinement region on $16^3\times 4$ lattice.
For $\beta <\beta_c(=2.298)$, a long loop exists in each configuration 
and all other loops are short.
Long loops have a characteristic length at each $\beta$ 
and the length becomes short around $\beta_c$.
For $\beta >\beta_c$, no long loop exists.
In $T\neq 0$ $SU(2)$ QCD, the monopole currents contained in a long loop 
can reproduce the full value of the string tension \cite{ejiri}.
This suggests that the long loop is related 
to the confinement mechanism in $T\neq 0$ QCD.

Next, we investigate the distribution of monopole currents 
contained in a long loop from its center.
The mean square of the distance from the center is calculated.
If the monopole currents exist uniformly in the $N_s^3\times N_t$ lattice, 
$R^2$ should be
\begin{equation}
   \frac{1}{N_s^3N_t}\int d^4x
      (x^2+y^2+z^2+t^2) 
   =\frac{N_s^2}{4}+\frac{N_t^2}{12}   .
\label{eqn:R}
\end{equation}
%
%---------- fig2 ---------
\begin{figure}[htb]
\vspace{3pt}
%\framebox[55mm]{\rule[-21mm]{0mm}{43mm}}
\epsfxsize=0.5\textwidth
\vspace{-27pt}
\begin{center}
\leavevmode
\epsfbox{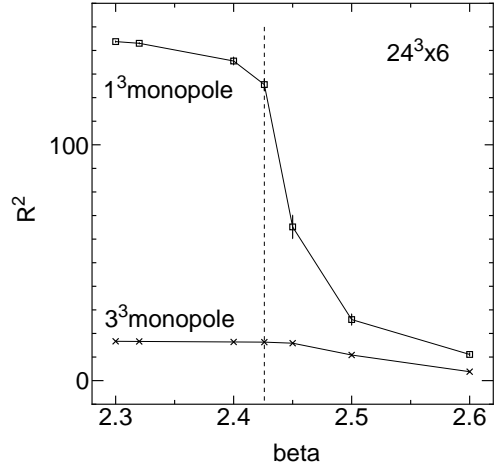}
\end{center}
\vspace{-50pt}
\caption{
$R^2$ versus $\beta$ on $24^3 \times 6$ lattices.
The dashed line indicates the location of the $\beta_c$.
}
\label{fig:distribution}
\vspace{-15pt}
\end{figure}
%-------------------------
%
$R^2$ versus $\beta$ on $24^3\times 6$ lattices is shown 
in Fig.\ref{fig:distribution}.
For $\beta <\beta_c$($=2.426$), the monopole loop is almost uniform 
in the whole lattice because the data are consistent with Eq.(\ref{eqn:R}).
For $\beta >\beta_c$, it is not uniform.
In the confinement phase, suppose that an $n^3$ monopole loop spreads 
uniformly through the lattice, we may define the effective size 
of monopoles as follows:
\begin{equation}
   l^3(n)=\frac{(N_s/n)^3\times (N_t/n)}
               {\langle L(n)\rangle} ,
\end{equation}
where the numerator is the volume of the effective lattice 
of an $n^3$ monopole, which we call renormalized lattice. 
The denominator is the length of the monopole loop in the 
renormalized lattice unit.
We define $\bar{l}= [l]+1$, 
where the symbol $[l]$
is the integer not exceeding $l$.
If the monopole current with a range $\bar{l}$ is set on a link, 
the nearest current can be put on links $\bar{l}$ apart 
in the renormalized lattice unit.
Hence for $n^3$ extended monopoles having the rounded range $\bar{l}$, 
the entropy can be discussed on a reduced lattice defined by 
\begin{equation}
   \left(\frac{N_s}{n\bar{l}}\right)^3\times 
   \left(\frac{N_t}{n\bar{l}}\right).
\end{equation}
%
%---------- fig3 ---------
\begin{figure}[htb]
\vspace{3pt}
%\framebox[55mm]{\rule[-21mm]{0mm}{43mm}}
\epsfxsize=0.5\textwidth
\vspace{-27pt}
\begin{center}
\leavevmode
\epsfbox{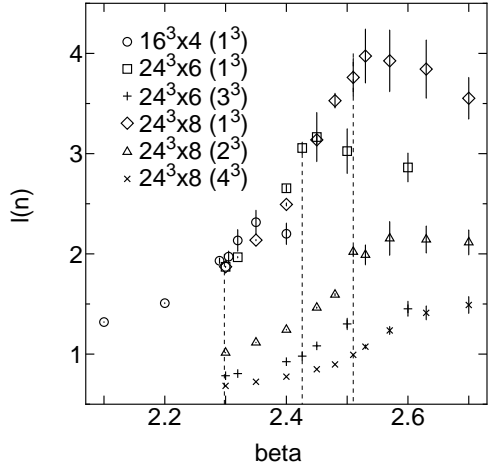}
\end{center}
\vspace{-50pt}
\caption{
$l(n)$ versus $\beta$ on various lattices.
The parenthesized number represents the extendedness of monopoles.
}
\label{fig:exclusionvol}
\vspace{-15pt}
\end{figure}
%-------------------------
%
%% We can derive the conclusion in Table[x].
%% %---------------- table --------------------
%% \begin{table*}[hbt]
%% % space before first and after last column: 1.5pc
%% % space between columns: 3.0pc (twice the above)
%% \setlength{\tabcolsep}{1.5pc}
%% % =====================================================
%% % adapted from TeX book, p. 241
%% \newlength{\digitwidth} \settowidth{\digitwidth}{\rm 0}
%% \catcode`?=\active \def?{\kern\digitwidth}
%% % =====================================================
%% \caption{$l(n)$ at $\beta_c$ and reduced lattices}
%% \label{tab:effluents}
%% \begin{tabular*}{\textwidth}{@{}l@{\extracolsep{\fill}}rrrr}
%% \hline
%% original lattice & extendedness  & renomalized lattice & $l$ 
%% & reduced lattice \\
%% \hline
%% $16^3\times 4$ & $1$ & $16^3\times 4$ & $2$ & $ 8^3\times 2$ \\
%% $24^3\times 6$ & $1$ & $24^3\times 6$ & $3$ & $ 8^3\times 2$ \\
%%                & $3$ & $ 8^3\times 2$ & $1$ & $ 8^3\times 2$ \\
%% $24^3\times 8$ & $1$ & $24^3\times 8$ & $4$ & $ 6^3\times 2$ \\
%%                & $2$ & $12^3\times 4$ & $2$ & $ 6^3\times 2$ \\
%%                & $4$ & $ 6^3\times 2$ & $1$ & $ 6^3\times 2$ \\
%% \hline
%% %% \multicolumn{5}{@{}p{120mm}}{Reprinted from: G.M. Ritcey,
%% %%                              Tailings Management,
%% %%                              Elsevier, Amsterdam, 1989, p. 635.}
%% \end{tabular*}
%% \end{table*}
%% %-------------------------------------------
%
Fig.\ref{fig:exclusionvol} shows a remarkable fact about $l(n)$ at $\beta_c$.
$l(n)$ is always equal to the half of the time extent 
of the renormalized lattice ($N_t/(2n)$) at $\beta_c$.
For example on $24^3\times 8$ lattice, 
$l(n)$ is equal to two at $\beta_c$ for $2^3$ monopole 
and equal to one for $4^3$ monopole.
Then the entropy of an $n^3$ extended monopole loop for any $n$ 
should be calculated always on reduced lattices with 
the time extent 2.
This means that the entropy of $n^3$ monopole is the same for any $n$ 
if the original lattice is the same.
What is the entropy of the monopole loop on the reduced lattice 
with the time extent 2?
It is apparent that the entropy is smaller than $\ln 7$ per unit length 
in the $T=0$ case.
The monopole loop is very long and occupies 
almost all the reduced lattice sites.
The time extent is only two and 
monopoles can not be regarded as completely point-like at $\beta_c$
due to their effective size.
The entropy is difficult to estimate accurately.
Anyway the entropy for any extended monopole on the same original 
lattice is the same.

\begin{figure}[htb]
\vspace{3pt}
%\framebox[55mm]{\rule[-21mm]{0mm}{43mm}}
\epsfxsize=0.5\textwidth
\vspace{-27pt}
\begin{center}
\leavevmode
\epsfbox{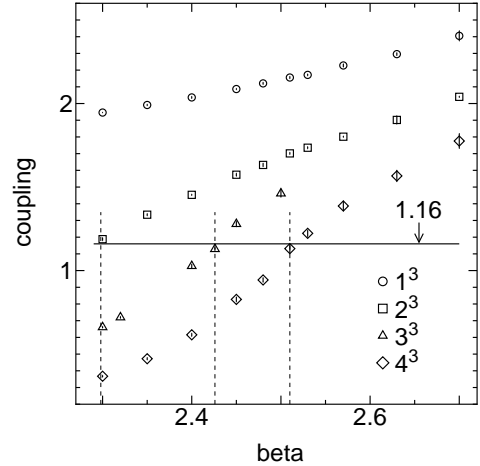}
\end{center}
\vspace{-50pt}
\caption{
The coupling constant of the dominant self-energy term 
of the monopole action versus $\beta$ 
for various sized monopoles 
on $24^3\times 8$ or $24^3\times 6$ lattices.
}
\label{fig:coupling}
\vspace{-15pt}
\end{figure}

On the other hand, we can see the energy per unit monopole loop length, 
which can be well approximated by the coupling constant 
of the self-energy term of the monopole action \cite{kita}.
The coupling constant is shown in Fig.\ref{fig:coupling}.
%
%---------- fig4 ---------
%-------------------------
%
We see that the energy is smaller for larger extended monopoles.
To determine the deconfinement transition point, 
larger monopoles may be more important.
What is the maximum size of extended monopole?
To define the monopole currents the renormalized lattice 
should have two independent lattice sites in the time direction.
This means the maximally extended monopole is always $(N_t/2)^3$ one.
The coupling constant of maximally extended monopoles takes about 1.16 
for each original lattice at $\beta_c$.
In Fig.\ref{fig:coupling}, 
$2^3$ ( $3^3$, $4^3$) monopoles are maximum for $N_t=4$ ( $6$, $8$) and 
the coupling constant is about $1.16$ at $\beta_c$,$2.298$ ( $2.426$, $2.51$).

In conclusion, 
if the entropy on a reduced lattice with the time extent 2 is proved 
to be about 1.16, the deconfinement transition is understood 
by balancing of the energy and the entropy of 
maximally extended monopole loops 
on any lattices.
It is interesting that the deconfinement transition may be explained 
by such a simple idea.
We should estimate the entropy accurately to prove the statement.

\end{document}